# An Overview of the Development of Stereotactic Body Radiation Therapy

Yanqi Zong[1,*], Zhengrong Cui[2], Luqi Lin[3], Sihao Wang[4], Yizhi Chen[5]

[1]tInformation Studies, Trine University, Phoenix, AZ, USA, yzong22@my.trine.edu

[2] Software Engineering, Northeastern University, Shanghai, China

[3] Software Engineering, Sun Yat-sen University, Shanghai, China

[4]Mathematics, Southern Methodist University, Dallas, TX, USA

[5] Information Studies, Trine University, Allen Park, MI, USA

* **Corresponding author**: Yanqi Zong (Email: zongyanqi@gmail.com)

**Abstract:** Stereotactic body radiation therapy (SBRT) refers to focusing high-energy rays in three-dimensional space on the tumor lesion area, reducing the dose received by surrounding normal tissues, which can effectively improve the local control rate of the tumor and reduce the probability of complications. With the comprehensive development of medical imaging, radiation biology and other disciplines, this less-fractional, high-dose radiotherapy method has been increasingly developed and applied in clinical practice. The background, radio-biological basis, key technologies and main equipment of SBRT are discussed, and its future development direction is prospected.

## 1. Introduction

Radiation therapy stands as an indispensable modality in cancer treatment, with as many as 70% of cancer patients necessitating some form of radiation therap [1]. Stereotactic radiotherapy technology, employing specialized radiation equipment to precisely focus high-energy rays on the target region, has emerged as a promising approach to cancer treatment. This technique minimizes complications associated with damage to normal tissue [2]. The evolution of linear accelerator technology and radiotherapy surgery has facilitated the application of stereotactic radiotherapy in the treatment of extracranial body tumors, surmounting challenges posed by tumor position variations due to respiration or movement [3]. Stereotactic body radiotherapy has gained widespread utilization across diverse cancer types, including lung, liver, kidney, pancreas, prostate, and spine tumors, resulting in favorable clinical outcomes [4,5].

The past two decades have witnessed a rapid development in stereotactic body radiotherapy, propelled by the comprehensive advancement of multiple disciplines such as radiation physics, radiation biology, clinical oncology, and medical imaging. Clinical studies, for example,[6] Liu using advanced machine learning with MRI data to detect early-stage Alzheimer's Disease and aims to create a predictive model for disease progression. reveal that hypofractionation therapy significantly enhances three-year local tumor control rates and survival rates in early non-small cell lung cancer (NSCLC) compared to conventional treatments. Additionally, it has demonstrated efficacy in treating rare metastatic lesions and primary lesions in the liver, kidney, pancreas, prostate, and spine. [7] Che study utilizes the Random Forest Tree method to enhance the classification of lengthy biomedical text documents related to cancer, addressing the challenge of processing research papers exceeding 6 pages in length.

Recent strides in artificial intelligence (AI), computer vision, and machine learning have also played pivotal roles in augmenting the precision and effectiveness of radiation therapy. AI and machine learning algorithms can predict tumor responses and evaluate treatment outcomes [8,9]. Concurrently, computer vision techniques enable real-time automatic identification and tracking of tumor targets during radiation therapy, enhancing precision and curtailing treatment duration[10].

The technical advancements in linear accelerator technology and radiation therapy surgery have greatly improved the precision and effectiveness of radiotherapy. The following section will provide an overview of some of the recent developments in these areas, with reference to several studies [11-24].

Moreover, the confluence of mechanical engineering and computer science has introduced notable innovations into the realm of radiation therapy.[25], for instance, conducted research at Dortmund University of Technology, delving into the synergistic blend of mechanical engineering and computer science to optimize logistics automation, with a focus on precise robot control achieved through accurate positional information.

Furthermore, [26] harnessed the power of the Xception model and data augmentation techniques for automating the quality inspection of casting product images, resulting in marked improvements in accuracy and efficiency. [27] introduced a project leveraging bio-inspired swarm intelligence for communication-free group object recognition, with a spotlight on neighbor observation.[28] delved into user preference analysis for smartphones and apps, emphasizing the significance of methods grounded in entity similarity and semantic assessment to gauge user satisfaction. Finally, [29] unveiled the Twins-PCPVT model for diabetic retinopathy (DR) detection, highlighting its prowess in capturing global and local features in fundus images, thereby enhancing diagnostic accuracy.

## 2. Basics of Radiation Biology

Compared with conventional radiotherapy methods, stereotactic body radiotherapy has the technical characteristics of less radiation times (1-5 times), high single-irradiation dose (8-30Gy), and higher accuracy and conformity of the target area. Such fractional treatment mode can increase the total dose and fractional dose of tumors, reduce the total dose and fractional dose of normal tissues, especially sensitive organs, and shorten the total treatment time. On the one hand, stereotactic body radiotherapy conforms to the general law of "effect-dose" in tumor radiobiology, directly or indirectly causing DNA double-strand breaks in tumor cells and killing tumor cells. On the other hand, high-dose irradiation will increase the damage to the tumor in other ways.

Currently, damage to the double helix structure of cellular DNA is considered to be the main cause of cell death. The classical radiobiological effects of ionizing radiation can be explained by a linear quadratic model (LQ model). Cell death by ionizing radiation has two separate components, one proportional to the dose and the other proportional to the square of the dose. The cell survival rate under the action of ionizing radiation can be expressed as formula 2-1.

$$S = e^{-\alpha \cdot D} - \varphi^{3 \cdot D^2} \qquad (2\text{-}1)$$

S is the fraction of cells surviving at dose D, and α and β are constants. The linear part is the result of two breaks from a single charged particle or photon, with probability of occurrence proportional to the dose irradiated. If two fractures are caused by differently charged particles, the probability of occurrence is proportional to the square of the irradiation dose, as shown in Figure 1. Therefore, high-dose irradiation can damage the DNA of tumor cells more effectively.

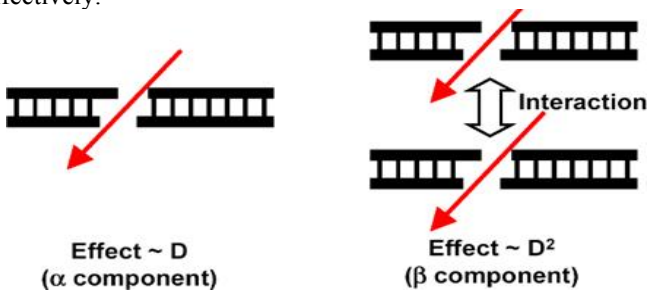

**Figure 2.1.** Schematic illustration of DNA strand breaks caused by ionizing radiation.

According to the different response time of tissues to radiation, it can be divided into early-response tissues and late-response tissues. Early response tissue has a fast renewal rate and a high α/β value, which is sensitive to the total treatment time, and shortening the total treatment time can aggravate the damage to tumor cells. Late-response tissues have slower turnover, lower α/β values, and are more sensitive to fractionated doses. Increasing fractional doses will aggravate the damage to late-response tissues. In addition, late responding tissues substantially did not proliferate during treatment, so changing the total treatment time had little effect on them. Malignant tumors are usually regarded as early-response tissues. One of the advantages of stereotactic body radiation therapy is that the total treatment time is short and it can effectively kill tumor cells.

## 3. Key Technology

Due to the technical characteristics of less fractionation and hypofractionation in stereotactic body radiotherapy, the single treatment time becomes longer and the single irradiation dose increases. Such an irradiation method will cause irreversible radiation damage to the normal tissue entering the treatment target area, that is, the late responding tissue. Therefore, the treatment equipment is required to fully guarantee the accuracy of clinical treatment, and the body stereotactic radiotherapy equipment is required to realize key technologies such as good respiratory management, real-time image-guided tracking treatment, accurate body position fixation and body position correction.

### 3.1. Respiratory Management

One of the major difficulties in the application of stereotactic radiotherapy to body tumors is to overcome the changes in the tumor location caused by the patient's physiological movements (heartbeat, breathing, blood sugar changes, etc.). Breathing has the most obvious effect on tumor location during physiological exercise. The management of respiratory motion can significantly reduce the boundaries of the planned target volume and reduce the damage to normal tissues. At present, the commonly used breathing management methods include: (1) Restricting breathing movement, that is, using certain technical means to reduce the breathing range or achieve the purpose of apnea, including abdominal pressurization technology, breath-holding technology after deep inhalation, and active breathing control technology, etc. . This approach has limited efficacy, can cause discomfort to the patient, and is poorly reproducible between treatments. (2) Respiratory gating technology, the most commonly used respiratory management scheme, refers to the combination of images collected by 4DCT and breathing depth through non-invasive markers, and the relative position of markers and tumors is used to correctly assess the tumor's degree of respiration within the respiratory cycle. Movement and changes in position relative to the treatment room. However, there is a difference in dynamic position between the marker and the tumor itself, and real-time imaging, such as real-time X-ray transmission, is required to monitor the tumor position and evaluate tumor mobility during treatment. (3) Real-time tumor tracking technology, that is, the application of various imaging techniques to monitor the location of the tumor in real time, and implement radiotherapy according to the location of the tumor at that time, so as to keep the relative position of the ray beam and the tumor unchanged, so as to reduce the impact of tumor movement on radiotherapy. The purpose of influence, this method requires higher equipment.

## 3.2. Fixing Position

In addition to the change of tumor position, the main factors affecting the boundary of the planned target area are setup errors. Choosing a position fixation technique with good repeatability is an important prerequisite for the implementation of stereotaxic body radiotherapy. Different from conventional fractionated radiotherapy, the position fixation technique of stereotactic radiotherapy not only needs to ensure the repeatability of the patient's position between treatment sessions, but also ensures that the patient's position remains unchanged within the treatment sessions due to the prolongation of treatment sessions. The current body position fixation technology includes stereotaxic frame fixation technology, implanting metal markers around the tumor, and frameless technology that directly uses the relative positional relationship between bony landmarks and the target area to implement online body position verification. It is worth noting that the fractional treatment time of stereotactic radiotherapy is significantly prolonged, and it is of great importance to choose a comfortable position and fixation device for the patient.

## 3.3. IGRT

Image guidance includes the following aspects: (1) image positioning of the target area; (2) real-time image tracking of the target area; (3) real-time correction of treatment plan (ART) under image guidance; (4) formulation of multimodal image treatment plan.

Stereotactic radiotherapy has a short treatment time and a large dose gradient between the tumor target area and surrounding normal tissues. Effective body position verification is required before treatment, and setup errors should be corrected in real time during treatment to ensure that the target area receives the prescribed dose and the surrounding normal tissues are spared or less exposed to radiation. From portal plain film, electronic portal imager (EPID) to slide CT technology, from megavoltage cone beam (MV CBCT) to kilovolt cone beam (KV CBCT) and then to digital tomosynthesis (DTS), the image quality of body position verification before treatment has been continuously improved, providing more abundant and accurate information. It enables doctors to better compare the images after setup and the planned images, discover and correct setup errors online, and increase the accuracy of treatment. CBCT is the mainstream IGRT technology at present. It adopts the principle of volume imaging and uses multiple two-dimensional images obtained by flat-panel detectors to generate three-dimensional images for volume image-guided radiotherapy. However, there are few research results on the application of body position verification techniques such as MV CBCT, KV CBCT and DTS in stereotactic radiotherapy. With the development of these imaging technologies, pre-treatment body position verification and real-time setup error correction (i.e. online corrected IGRT technology) can be more accurately realized, further reducing the planned target area and improving treatment accuracy.

Adaptive radiation therapy, i.e. IGRT technique with offline correction. Yan first proposed the concept of adaptive radiation therapy in 1997, that is, observing changes in patients during the first few treatments, including changes in position and changes in dose-response relationships, etc., and comprehensively improving radiotherapy plans based on these changes. At present, in practice, adaptive radiotherapy mainly considers the change of the position. During the first few (5-9) treatments, the two-dimensional or three-dimensional images of the patient are collected, and the set-up error of each time is measured offline to predict the set-up of the entire course of treatment. error, and then adjust the treatment plan accordingly.

Multimodal imaging treatment plan formulation refers to the integration of the advantages of multiple imaging methods, the combination of structural imaging and functional imaging, and the delineation of tumor target areas and the formulation of radiotherapy plans. Commonly used multimodal imaging includes CT/PET, CT/SPECT, CT/MRI, and PET/MR that have appeared in recent years. Multimodal imaging technology has significant advantages over traditional single-modal imaging technology, which can provide richer structural and functional information, taking into account tumor heterogeneity and metabolic information. However, image registration is difficult and the workload of image fusion is heavy. At present, there is not enough evidence to prove that multimodal imaging technology can improve the effect of cancer radiotherapy.

## 3.4. Static/Dynanmic IMRT

Hypofractionated and less fractionated body stereotactic radiotherapy requires higher target volume conformity and more appropriate dose distribution within the radiation field. The implementation of conformal IMRT relies on multi-page collimation gratings (MLC). A complex shape of the radiation field is formed by a certain number of blade movements, and a radiation field that meets the treatment needs can be obtained by changing the shape of the radiation field and modulating the distribution of the dose within the radiation field range. Its important parameters are: (1) Leakage between blades; (2) Blade thickness; (3) Blade over-middle distance; (4) Blade moving speed; (5) Penumbra size; (6) Blade positioning accuracy, etc. Static intensity modulation means that the field shape does not change during the irradiation process, but changes in the middle of adjacent fields, also known as Step and Shoot. The gantry angle remains unchanged during the irradiation process, and different weights of each subfield are added to obtain the desired intensity distribution. The conformity of the static intensity modulation is insufficient, and the treatment time is long, so the dynamic intensity modulation technology was developed. The shape of the grating changes continuously during the irradiation process, also known as the sliding window (Sliding Window). The velocity of each pair of blades during irradiation is a function of time. The equipment is required to have the dose rate servo function and the precise control capability of the position of the grating blades, and the dynamic intensity modulation dose field is more delicate and has a higher degree of conformity.

## 3.5. FFF

In order to increase the dose rate of body stereotactic radiotherapy and reduce its treatment time, the FFF system has been applied in the main body stereotactic radiotherapy equipment. Studies have shown that removing the leveler can effectively increase the dose rate, reduce scattering and leakage, reduce the penumbra, and reduce the dose outside the radiation field. Can effectively kill tumor cells while protecting the surrounding normal tissue. However, FFF also has disadvantages. The leveler can filter some low-energy leakage electrons and photons. The energy spectrum of the ray shifts to the left, the low-energy components increase, the ray becomes softer, and the dose on the skin surface increases.

### 3.6. Non-conplanar and dynamic are illumination methods

Traditional three-dimensional conformal intensity-modulated radiation therapy adopts coplanar radiation fields. Such a radiation pattern will form a "high-dose shell" in the target area, and the dose of fractionated body stereotactic radiotherapy will increase significantly. Entering the "high-dose shell" normal The radiation damage to the tissue is obviously aggravated, so 6-10 non-coplanar radiation fields or dynamic arc irradiation methods are used to reduce the range of "high-dose shell", such as Gamma Knife and Cyberknife, using small beams , to complete the treatment of irregular target areas by means of scanning. However, the use of non-coplanar radiation fields will increase the collision probability between the accelerator head and the treatment bed. In addition, as the number of radiation beams increases, the possibility of radiation beams directly passing through key organs increases, which requires higher equipment requirements.

## 4. Development Tendency

In terms of radiobiology, it is necessary to further understand the radiobiological effects of high-dose and low-fraction irradiation on tumors, and to summarize a more suitable dose-effect model for SBRT to guide the formulation of radiotherapy plans. The systemic therapeutic effect of SBRT combined with chemotherapy, molecular targeted drugs, and immunotherapy is a research hotspot in cancer.

In terms of image guidance, real-time four-dimensional radiation therapy (R4RT) can minimize the uncertainty of tumor motion, which is the main development direction. In addition, the application of multimodal image fusion technology to target delineation is also a research hotspot. Through computer vision and big data , optimizing the image registration and fusion algorithm can make multimodal image fusion technology play a practical role in stereotactic radiotherapy.

In terms of the irradiation field segmentation algorithm, considering the basis of biological factors, it is urgent to break through a new mathematical method for selecting large-segmentation or multi-segmentation irradiation schemes based on physical dose distribution.

In terms of mechanical design, multiple robotic arms can be considered to work together to reduce the time of radiotherapy and reduce the pain of patients. Efforts are made to improve the movement accuracy of each component, so that the overall treatment accuracy of the equipment can reach sub-millimeter level.

In terms of acceleration tube design, the pursuit of miniaturization (C-band, X-band), reduce the weight of the whole machine, and reduce costs. Strive to increase the dose rate and maintain stability, to achieve adjustable energy, including MV/KV switching, to achieve homologous dual beams. We can make efforts to optimize the design and improve the behavior of crucial components of accelerators to finally improve the performance of medical accelerators. For example, X-ray target and corresponding cooling system can be optimized to improve the X-ray yield. New accelerating structure can be investigated to make the accelerators more compact and occupy little room. Other components including pulse compressor and electron gun can be optimized to improve the stability and efficiency of the accelerators. Explorations and experiments of different band accelerators should be performed to achieve better performance. In-depth physics study on beam transmission should be carried out to explore some new optimization methods.

The future of radiation therapy will require the interdisciplinary integration of imaging, physics, biology and oncology. We term this concept "dose-combined radiation therapy (DCRT)" and anticipate that future generations of radiation therapy will routinely incorporate these principles into clinical practice.